\documentclass[english,draft]{article}
\usepackage[T1]{fontenc}
\usepackage[latin9]{inputenc}
\usepackage{amssymb}
\usepackage{setspace}
\makeatletter
\newcommand{\lyxaddress}[1]{
\par {\raggedright #1
\vspace{1.4em}
\noindent\par}
}

\makeatother

\usepackage{babel}
\begin{document}

\title{Comments on ``There is no axiomatic system for the quantum theory''}

\author{J. Acacio de Barros}

\maketitle

\lyxaddress{Liberal Studies Program, 1600 Holloway Ave. BH 238, San Francisco
State University, San Francisco, CA 94132}
\begin{abstract}
In a recent paper, Nagata \cite{nagata2009thereis} claims to derive
inconsistencies from quantum mechanics. In this paper, we show that
the inconsistencies do not come from quantum mechanics, but from extra
assumptions about the reality of observables. 
\end{abstract}
\pagebreak{}

Quantum mechanics is one of the best tested theories in modern physics,
and yet there is no consensus as to what it means. The reason lies
in the fact that, as Feynman eloquently put, ``nobody understands
quantum mechanics'' \cite{feynman1965lectures}. This lack of understanding
comes from the difficulties to interpret even the simplest of the
examples, like the two-slit experiment, in ways that are consistent
with the observations and with an underlying ontology that most consider
satisfactory (for distinct approaches, see \cite{suppes1994diffraction,suppes1996photons,omnes1994theinterpretation,holland1995thequantum,suppes1996violation,de2007realism}).
For example, quantum mechanical observables do not allow for standard
joint probability measures to be defined \cite{de_barros_probabilistic_2010,suppes1981whenare,de2001probabilistic,de2000some},
and if we assume such probabilities, we derive contradictions \cite{barros2000inequalities}.
Furthermore, the structure of observables does not satisfy a classical
logic, but instead a quantum one \cite{pitowski1989quantum}. Thus,
an area of intense interest in the foundations of quantum mechanics
is the search for theories that complete quantum mechanics, such as
hidden-variable theories, and give sense to it \cite{holland1995thequantum}. 

In a recent paper, Koji Nagata looked into the possibility that quantum
mechanics leads to contradictions \cite{nagata2009thereis}. Though,
as mentioned in many of the references above, contradictions can be
derived depending on the assumptions used, Nagata goes further and
claim that ``there is a contradiction within the Hilbert space formalism
of the quantum theory.'' He then concludes that no axiomatization
exists for quantum mechanics. In this paper, we clarify some assumptions
made by Nagata and show that the derived contradictions are not part
of the theoretical structure of the theory, but instead are part of
metaphysical assumptions about the systems. Therefore, the contradictions
obtained by Nagata are not an impediment to the axiomatization of
the theory, but instead to a specific worldview. 

Let us start with Nagata's derivation of a contradiction. In \cite{nagata2009thereis},
a pure state  spin-$\frac{1}{2}$ system on the $x$-$y$ plane is
considered. He then goes on to show that if we compute the quantum
mechanical expectation of such state measured in an arbitrary direction
$\vec{n}$, the expected value $E_{QM}\leq1$, which implies that
$\left|E_{QM}\right|_{\max}=1$. This result is consistent with the
fact that, for his choice of units $\hbar/2=1$, if the system is
prepared in the same direction as $\vec{n}$, we always get the same
answer as $1$. Finally, Nagata shows that if we use a collapsed state
and then compute the expectation, $E'_{QM}\leq2$, implying $\left|E'_{QM}\right|_{\max}=2$
(we changed the notation to $E'$ to avoid confusion with the previous
value). Thus, Nagata claims, because $\left|E_{QM}\right|_{\max}$
cannot have two different values, we arrive at a contradiction. Before
we proceed, we would like to point out that Nagata's inequalities
do not by themselves imply a contradiction. For example, the statements
$b\leq1$ and $b\leq2$ are not contradictory, as $b=0$ is an example
that satisfies them. To prove a contradiction, Nagata would have to
construct a system which he could prove not only is less than $2$
but is also greater than $1$. Though he did not show such proof,
it in fact exists. But to make clear where the contradiction comes
from, we present it below in a simplified version. 

At the core of Nagata's derivation lies an important feature of quantum
mechanics, namely that if you do not measure something you cannot
assume that it has a value. In other words, assuming values to unmeasured
observables leads to contradictions (see \cite{suppes1998acollection,barros2000inequalities}
and references therein). Let us analyze the case of a spin-$\frac{1}{2}$
system. First, let us see what quantum mechanics can tell us about
this system. If we want to observe its spin in a given direction $\vec{m}$,
the associated observable is $\hat{O}_{\vec{m}}\equiv\vec{m}\cdot\vec{\sigma}$,
where $\vec{\sigma}$ is a vector in $\mathbb{R}^{3}$ with the Pauli
matrices as components, i.e. $\vec{\sigma}=(\hat{\sigma}_{x},\hat{\sigma}_{y},\hat{\sigma}_{z})$.
From the properties of the Pauli matrices, it is easy to show that
$\hat{O}_{\vec{m}}$ has eigenvalues $\pm1$, regardless of the measurement
direction. Since $\vec{m}$ is arbitrary, let us we pick three distinct
directions, $\vec{e}_{1}$,$\vec{e}_{2}$, and $\vec{e}_{3}$, such
that $\vec{e}_{1}+\vec{e}_{2}+\vec{e}_{3}=0$. The corresponding observables
will be $\hat{O}_{1}\equiv\vec{e}_{1}\cdot\vec{\sigma}$, $\hat{O}_{2}\equiv\vec{e}_{2}\cdot\vec{\sigma}$,
and $\hat{O}_{3}\equiv\vec{e}_{3}\cdot\vec{\sigma}$. Quantum mechanics
not only tells us that measuring $\hat{O}_{1}$, $\hat{O}_{2}$, or
$\hat{O}_{3}$ yields $\pm1$ values, but it also tells us that we\emph{
cannot measure them simultaneously, }as they do not commute. 

A natural question to ask is the following. Is it possible to assign
a value to spin, even though a measurement has not been performed?
To answer this, let us assume that we indeed can assign such value
(we follow \cite[pp. 15-16]{peres1993quantum}). Let $\mathbf{P}$
be a vector random variable corresponding to the actual value of the
system's spin before any measurement. It follows that if we measure
it in a direction $\vec{m}$, the outcome of the experiment must be
$\vec{m}\cdot\mathbf{P}$. Now, quantum mechanics tells us that, regardless
of the direction, $\vec{m}\cdot\mathbf{P}$ will take values $+1$
or $-1$. But, using the vectors we picked before, we have 
\begin{eqnarray}
\vec{e}_{1}\cdot\mathbf{P}+\vec{e}_{2}\cdot\mathbf{P}+\vec{e}_{3}\cdot\mathbf{P} & = & \left(\vec{e}_{1}+\vec{e}_{2}+\vec{e}_{3}\right)\cdot\mathbf{P}\label{eq:contradiction-1}\\
 & = & 0.\nonumber 
\end{eqnarray}
This, of course, leads to a contradiction, as the sum of three $\pm1$
random variables cannot equal zero. 

An analysis of the above example shows the origin of the contradiction.
Since quantum mechanics forbids the simultaneous measurements of $\hat{O}_{k}$,
as they do not commute, it does not allow us to simultaneously assign
values to them. The contradiction does not come from quantum mechanics,
but from the assumption that we can assign values to measurements
that were not performed. But not even assigning values leads to contradiction,
if we are careful. For example, we could assign values to $\vec{e}_{1}\cdot\mathbf{P}$
and to $\vec{e}_{2}\cdot\mathbf{P}$, as long as we assumed that the
$\mathbf{P}$ in $\vec{e}_{1}\cdot\mathbf{P}$ is different from the
one in $\vec{e}_{2}\cdot\mathbf{P}$, a feature called contextuality
\cite{despagnat1999conceptual}.

The above example contains the essence of Nagata's argument. By computing
the value of a quantity using the quantum mechanical formalism yields
different quantities than by computing it from the distribution over
the random variables associated with the (non-commuting) observables
in quantum mechanics (his use of values from von Neuman's projections).
The reason for this discrepancy in computations is that, in the latter
case, there is an underlying assumption that an unmeasured quantity
exists \emph{independent }of the other quantities. This, of course,
is not true, as quantum mechanical variables are contextually dependent
from each other, a characteristic stressed by Bohr. We emphasize that
this characteristic of quantum mechanics is not at all disturbing,
as it is common to many classical systems \cite{barros2009quantum}.
The troubling characteristic comes from a combination of contextuality
and non-locality, made famous by the Einstein-Podolsky-Rosen paper
\cite{einstein1935canquantum} and by Bell's inequalities \cite{bell1966onthe}. 

Quantum mechanics is indeed a strange theory. But its strangeness
comes not from an inconsistency of its mathematical structure, but
from the metaphysical views it imposes on us. If we insist on having
worldviews where the values of variables exist independent of the
observation, then we will get into contradictions. But, as many authors
show, such contradictions can be avoided by carefully interpreting
the meaning of the mathematical formalism. 

The author would like to thank Dr. Gary Oas for his comments on the
final version of the manuscript. 

\bibliographystyle{/Users/barros/Library/texmf/bibtex/spphys}
\bibliography{MainBibliography,/Users/barros/Desktop/Work/Research/VaxjoConference/VaxjoProbPaper,citations}

\end{document}